\documentclass[default]{aastex701}

\begin{document}

\title{Ranking the drivers of the Venusian bow shock and ion composition boundary locations}

\author[orcid=0000-0002-9683-9657,gname=Philippe,sname=Garnier]{P. Garnier}
\affiliation{IRAP, Université de Toulouse, CNRS, CNES, Toulouse, France}
\email[show]{philippe.garnier@utoulouse.fr} 

\author[orcid=0000-0003-3497-3209,gname=Moa,sname=Persson]{M. Persson}
\affiliation{Swedish Institute of Space Physics, Uppsala, Sweden}
\email[show]{moa.persson@irf.se} 

\author[orcid=0000-0003-0673-2264,gname=Sofia,sname=Bergman]{S. Bergman}
\affiliation{KTH Royal Institute of Technology, Stockholm, Sweden}
\email[show]{sobergm@kth.se} 

\author[orcid=0000-0002-7056-3517,gname=Yoshifumi,sname=Futaana]{Y. Futaana}
\affiliation{Swedish Institute of Space Physics, Kiruna, Sweden}
\email[show]{futaana@irf.se}

\begin{abstract}

\par The Venusian interaction with the solar wind leads to the formation of an induced magnetosphere structured by plasma boundaries. Their dynamics is complex, due to the combined influence of external (solar photons, solar wind plasma and interplanetary magnetic field (IMF)) and internal (ionized atmosphere) drivers. Studying these drivers helps understanding the transfer of energy and momentum throughout the Venusian system, and has thus implications for the erosion of the atmosphere through its coupling with the solar wind.
\par We here analyze and rank the influence of the main drivers of the Venusian bow shock and ion composition boundary locations. We revisit the results by \citet{Signoles_2023} based on Venus Express measurements by combining several methods such as the Akaike Information Criterion, Least Absolute Shrinkage Selection Operator regression, and partial correlations. These methods allow to investigate cross correlations that appear and can bias the interpretation, and allow to rank drivers with robust approaches.
\par The bow shock appears primarily driven by the IMF intensity or Mach number, the IMF $\theta_{bn}$ angle separating quasi-perpendicular vs quasi-parallel shocks, and then the solar extreme ultraviolet fluxes and solar wind dynamic pressure (with little influence of the convective electric field induced asymmetries). The Ion Composition Boundary is primarily driven by extreme ultraviolet fluxes, with a more reduced influence of several solar wind parameters and IMF induced magnetic pileup asymmetries. We also compare the behaviors of both boundaries and then compare the bow shock driver rankings at Mars and Venus. Finally we propose an analysis of the drivers of the extreme bow shock and ion composition boundary excursions.

\end{abstract}



\section{Introduction}

\par Venus interacts continuously with the dynamic solar wind (SW), inducing a number of physical processes at local and global scales. The interaction is determined by the characteristics of both the incoming plasma flow (supersonic and superalfvenic, with an associated interplanetary magnetic field (IMF)) and of the planet itself (no intrinsic magnetic field, thick atmosphere). At a global scale, an induced magnetosphere is thus created, structured by plasma boundaries that separate various regions and whose geometry and dynamics shape the global interaction of the planet with the upstream SW \citep{luhmann_j:86, futaana_y:17}. The dynamics of these plasma boundaries have important implications for the transfer of energy and momentum throughout the Venusian system, and thus for the erosion of the atmosphere through its coupling with the SW \citep{Persson_m:21}. Therefore, analyzing the dynamics, shape, and dependence of these boundaries on external forcing is crucial for understanding the atmospheric evolution of Venus.

\par The Venusian bow shock (BS) is a fast magnetosonic wave standing in the SW, that slows down and heats the SW plasma. The variations of this boundary were first investigated by the Pioneer Venus Orbiter mission in 1978-1992 \citep{colin_l:80}. Several studies suggested that the shock drivers include magnetosonic Mach number, SW dynamic pressure, solar cycle, and IMF orientation \citep{alexander_russell:85, zhang_tl:90}. In particular, it was found that the IMF orientation and strength have an impact on the BS asymmetries in the VSE frame (Venus-Solar-Electric field coordinates, with x axis pointing to the Sun and z axis along the upstream SW motional electric field, and y axis completing the orthogonal system). The BS showed asymmetries between the magnetic pole vs. equator regions, between dawn and dusk, and between the north and south magnetic polar regions. These asymmetries were interpreted as a result of anisotropies of the magnetosonic wave velocity or the strong mass loading by pickup ions, which is also dependent on the cone angle and its effect on the convective electric field $\overrightarrow{E}= - \overrightarrow{V} \times \overrightarrow{B}$ \citep{russell_ct:88, khurana_kk:94, alexander_cj:86}.

\par The Venus Express mission \citep{svedhem_h:07} then allowed for revisiting our knowledge of the Venusian plasma boundaries. The asymmetries of the BS were confirmed, and were attributed to either the influence of the tangential (to the BS surface) component of the IMF \citep{chai_l:14}, or to the anisotropic magnetosonic wave speed \citep{chai_l:15}. \citet{Signoles_2023} used the full Venus Express data set (2006-2014) and concluded that the BS location was largely dependent on the solar cycle and the SW dynamic pressure, in combination with the aforementioned asymmetries and their association with asymmetric pickup ion distributions and mass loading effects. These authors also investigated the Ion Composition Boundary (ICB) - which separates the energetic SW plasma from ions of planetary origin \citep{holmberg_m:17} - and suggested that the ICB on the dayside is compressed by the increased dynamic pressure and IMF intensity during solar maximum (2011–2014), but to a smaller or insignificant degree during solar minimum (2006–2011) conditions.

\par Recently, efforts have been made to create multiparametric fits to the boundary locations. \citet{Wang_2024_a} used simulations from a 3D magnetohydrodynamic (MHD) model to determine parametric relations that could reproduce the influence of the IMF intensity and the magnetosonic Mach number (Mms). The results indicated a significant influence of both the Mms and IMF components perpendicular to the boundary surface on the BS position and geometry, as a result of the propagation of magnetosonic waves. \citet{Wang_2024_b} performed a statistical study of Venus Express data to build an empirical model of the BS, where they included both the influence of the Mms and EUV radiation flux. They left out the IMF intensity, IMF cone angle and SW dynamic pressure from the parametric 3D model as there were no clear correlations between these parameters and the BS, and did not include the IMF $\theta_{bn}$ angle (between the normal of the BS and the IMF vector) despite its apparent influence on the BS shape. 

\par Similarly, \citet{Rollero_2025} used fit functions of the semi-latus rectum (L) for the BS and the radius of the ICB (assumed circular on the dayside) and determined the main drivers based on the best linear regressions. They chose sets of drivers based on best regressions and incorporated them into parametric models: proton flux (instead of dynamic pressure and estimated from a solar wind propagation model), IMF intensity and $\theta_{bn}$ angle for the BS (not EUV fluxes) ; proton flux and EUV fluxes for the ICB.

\par In summary, previous studies have conducted analysis of the boundaries and their variations with external drivers using direct data analysis, as well as physical and parametric modeling. However, their methods often rely on direct analysis of scatter plots and correlation factors, confined to analyzing a single or a few parameters at a time. In reality, the dynamics of these boundaries are complex, and multiple drivers may influence the boundaries simultaneously at different orders of magnitudes with cross-correlations between driver parameters that may significantly affect the results. Cross-correlations between parameters are often tackled by separating the datasets into subsets, where one parameter is controlled, which can lead to a significantly reduced dataset where sampling biases are introduced. Additionally, smaller drivers are challenging to separate out through these methods due to the typically limited number of measurements available. Moreover, the difference in mission, time periods and drivers included in every study may lead to contradictory conclusions. For example, the EUV flux was considered a major driver for both the BS and ICB for most studies \citep{Wang_2024_a, russell_ct:88, alexander_russell:85}, while others did not find it important for the boundary location \citep{Rollero_2025, Signoles_2023}. In a similar manner, SW dynamic pressure is usually considered important for the BS except for some authors (e.g. \citet{Wang_2024_b}).

\par In order to mitigate these issues, \citet{garnier_p:22a} and \citet{garnier_p:22b} (hereafter G22) proposed an alternative methodological approach to help quantify and rank the relative influence of possible drivers of plasma boundaries. They showed that in a complex system, such as the Martian BS dynamics, a combination of several methods adapted to complex correlated systems can extract the relative importance of a large set of drivers. Notably, they showed that the cross correlations between the expected parameters, such as Mms, IMF or SW dynamic pressure, or unexpected, such as crustal magnetic fields and EUV flux, can lead to inaccurate conclusions when using the common single parametric investigations. For example, G22 showed that the crustal field and $\theta_{bn}$ angle are significant drivers, and are typically underestimated in boundary studies using the Mars Express and MAVEN datasets.

\par In this paper, we propose to revisit the work by \citet{Signoles_2023} (hereafter S23) on the drivers of both the ICB and BS locations, by using the same dataset but following a similar approach to \citet{garnier_p:22b}. Section \ref{DatasetMethod} describes the dataset and methods used for this study. Section \ref{Results} provides the results for the BS and ICB boundaries analysis, before an in-depth discussion in section \ref{Discussion} and conclusions in section \ref{Conclusions}.

\section{Datasets and methods}
\label{DatasetMethod}

\subsection{Datasets used}
\label{Dataset}

\par In this study, we investigate the drivers of the BS and ICB by using the boundary crossing dataset \citep{persson_2023_database} for Venus Express used by S23 and \citet{Rollero_2025}. We refer to S23 for a detailed description of the dataset, the upstream conditions and the spatial coverage of the dataset, as well as for the boundaries identification method, while a short summary is provided below.

\par Combined plasma and magnetic field measurements from the VEX mission were used to construct the crossings dataset. S23 used the Ion Mass Analyser (IMA) and Electron Spectrometer (ELS) of the Analyser of Space Plasma and Energetic Neutral Atoms (ASPERA-4 ; \citet{barabash_s:07}) for plasma measurements, as well as magnetometer (MAG ; \citep{zhang_t:06}) data. IMA measured ions at a cadence of $192$ $s$ in the energy range $0.01–36$ $keV$, with an energy resolution of $7\%$ and a field of view (FoV) of $90^{\circ}$ x $360^{\circ}$. The ELS sensor measured electrons at a cadence of $4$ $s$ with an energy resolution of $8\%$ in the energy range of $0.01–15$ $keV$, with a FoV of $5^{\circ}$ x $360^{\circ}$. The MAG instrument is a fluxgate magnetometer that measured the three magnetic field vector components with a resolution of up to 32 Hz, downsized to a resolution of $4$ $s$ in the dataset.
\par Visual inspection was performed to define the dataset \citep{persson_2023_database}, based on the following criteria. Inbound shock crossings were identified as sharp increases in the magnetic field total magnitude, accompanied by increases in the energetic ion and electron counts and temperatures (outbound crossings correspond to decreases). Inbound ICB crossings were identified from decreases in magnetosheath protons and electrons and increases in lower energy planetary heavy ions. Only dayside ICB crossings were included due to larger uncertainties on the nightside. This lead to a total dataset of 5193 BS crossings and 2679 ICB crossings.

\par In this study, we use a one dimensional approach, where we assume that the ICB can be described by the distance $\rho$ (i.e., circular shape on the dayside, as described by e.g. \citet{martinez_c:08}) and that the BS can be described by the extrapolated terminator distance $R_{TD}$ calculated from conic section fitting of the boundary, which is a common approach for the Venusian BS \citep{slavin_ja:80, martinez_c:08}. The $R_{TD}$ approach allows us to investigate the variability of the boundaries by removing the strong solar zenith angle influence, assuming an axisymmetric symmetry, and provides a method to investigate the presence of any large scale and permanent influence on the boundaries location, as performed in a number of previous studies (e.g. \citet{vignes_d:02, edberg_ne:09, chai_l:15, wang_m:20b}; G22).

\par For the BS, the crossings location is first transformed into the SW aberrated Venus Solar Orbital (VSO) coordinate system: in the non aberrated VSO frame, X points from Venus’ center toward the Sun, Y points in the Venus anti-orbital direction, and Z completes the righthand, while the aberrated frame includes a correction accounting for a constant $5^{\circ}$ aberration
angle about the Z-axis toward -Y. Then, $R_{TD}$ is given by:
\begin{equation}
R_{TD} = \sqrt{L^2+(e^2-1)\cdot X_0^2+2\cdot e\cdot L\cdot X_0}
\label{eq1}
\end{equation}

The focus of the conic is located at $(X_0,0,0)$ with $X_0 = 0.688$ $R_V$ (from S23) and $R_V=6052$ $km$ is the Venus radius. We used eccentricity $e$ values from S23 of respectively $e=1.042$ for solar minimum conditions (from 2006 to 2010) and $e=1.052$ for solar maximum conditions (from $1^{st}$ January 2011 to the latest 2014 measurements). The semi-latus rectum $L$ is recalculated from the knowledge of the angle between the $X$ axis and the crossing location before being used in equation \ref{eq1}.
\par The possible drivers considered are solar wind dynamic pressure, IMF strength and orientation, Alfven Mach number and sunspot number (SSN) as a proxy for EUV fluxes. SSN data were retrieved from \citet{SILSO_Sunspot_Number}. Upstream SW conditions were estimated from Venus Express ASPERA-4 IMA and MAG measurements in the upstream SW, and only used if stable enough conditions were encountered (adapted from S23): we selected data with less than $10$ $cm^{-3}$ and $70$ $km/s$ density and velocity variation over a 40 min time range around the BS crossings, as well as with IMF conditions from 20 minutes before (with an average inbound IMF intensity $B_i$) to 20 minutes after the BS crossings (with an average outbound IMF intensity $B_o$) such that $\frac{\mid | \overrightarrow{B_o} \mid - \mid \overrightarrow{B_i} \mid \mid}{| \overrightarrow{B_o} \mid + \mid \overrightarrow{B_i} \mid}<1 $ and less than $120^{\circ}$ between these vectors.

\par The selection criteria used to determine upstream parameters lead to a reduced dataset with all parameters available of $1604$ shock crossings and $916$ ICB crossings.
\par We note that the ion temperature from IMA was not considered in this study as the instrument was not designed to study SW ions \citep{barabash_s:06}, thus leading to unreliable temperature measurements at this stage. Thus, neither S23 nor this study could investigate the Mms, and instead rely on the Alfven Mach number M$_A$. \citet{Wang_2024_b}) and \citet{Rollero_2025} used either the calculated ion temperature moments or propagated ion moments, but the error estimation of using such data is out of scope for this study. The effect of these points will be discussed further in the discussion section.

\subsection{Methods}
\label{Methods}

\par We describe in this section the specific methods used in our paper to analyze the influence of the BS and ICB location drivers: correlations and partial correlations, Akaike Information Criterion, and Least Absolute Shrinkage Selection Operator. All calculations were performed with dedicated routines and libraries in the statistics R software.

\subsubsection{Correlations and partial correlations}

\par After a direct analysis of the BS extrapolated terminator distance or of the ICB distance with respect to parameters of influence, we will first show the cross correlations that may exist between parameters, based on the simple linear Pearson correlation coefficients (considered significant only with p-values below $5\%$, i.e. inside $2$ standard deviations for a gaussian distribution). 
\par Then, we will use and compare with a partial correlation approach to analyze the impact of possible cross correlations between parameters. This approach (see G22 or \citet{baba_k:04}) also assumes linear relations (or power laws linearized with a logarithm) and provides correlation coefficients and p values between two variables $y$ (i.e $R_{TD}$ or $\rho$) and $x_0$ (sunspot number, IMF intensity...), after controlling for the influence of others ($x_i$). The calculation of the partial correlations can be performed for any number of variables of influence and control, through a recursive approach. P values associated with these partial correlations are also provided based on a t-test to compare with the null assumption. Let us remind however that correlations cannot be used to determine rigorously rankings of drivers, on the contrary to dedicated methods such as AIC defined in the next subsection.

\par \citet{garnier_p:22a} managed to disentangle with partial correlations the impact of magnetic crustal fields on the Martian BS $R_{TD}$ distance despite a significant cross correlation with EUV fluxes in the MAVEN dataset. This method avoids using sub-selections of the dataset to control for specific variables, thus keeping all the information inside the dataset. Note that the linear hypothesis is weak and the method does not need true linear relationships to remain valid, since at first order most of the regular relationships can be considered as linear (or power law like).

\subsubsection{The Akaike Information Criterion}

\par Correlations are often used to compare the influence of various drivers, but correlations are only aimed at confirming statistical significance of relationships between parameters. On the contrary, several methods are designed to rank models or drivers, such as the Akaike Information Criterion (AIC) or its variant the Bayesian Information Criterion (BIC), or to some extent regularization techniques such as LASSO described below. We refer to Appendix A and B in G22 for more details on these techniques.
\par The Akaike information criterion is a model selection approach developed in the frame of information theory, that can rank statistical models compared to a real dataset \citep{akaike_h:74}. AIC provides a score for each model: $\mathrm {AIC} \,=\,2k-2\ln({\hat {L}})$ with k number of parameters of each model and $\hat L$ maximum value of the likelihood function. Calculating the exponential of the difference between the individual scores of two statistical models provides the relative likelihood for the ranking proposed between them. 

\par In our case, we compare the full dataset including all variables with alternative models corresponding to linear (or power laws then linearized) combinations of variables where, one by one, each variable is removed. The algorithm - stepwise backwards regression - then provides scores that represent the amount of information lost by each model (each corresponding to one missing variable), ultimately providing a ranking of the most influencing variables (i.e. those inducing the largest loss of information when removed). The $2k$ term is a penalty term that prevents overfitting by worsening the score when more variables are included in the models. The BIC approach - $\mathrm {BIC} \,=\,k ln(n)-2\ln({\hat {L}})$ with n number of observations - can be used to increase the penalty term. Note that AIC or BIC only provide a relative quality of each model versus others, but do not provide a quality level of these. Thus, our routine also includes intermediate steps to remove model variables with too low direct correlation significance (p values $>0.1$) or too strong multicollinearity levels with other variables (i.e. Variance Inflation Factor values above 4).

\subsection{Lasso regularization}

\par The Least Absolute Shrinkage Selection Operator (LASSO) is a commonly used supervised L1 regularization method for regressions \citep{tibshirani_r:96}, that helps identifying the most significant predictors in a regression model by reducing the dimension of the problem. It is similar to classical multivariable linear regression, but with a penalty over the sum of the absolute values of the regression coefficients to limit overfitting and select influent variables. One assumes a parameter of interest $y$ depends at first order on a linear combination of variables $x_i$ with an error term $e$:
\begin{equation}
y= \sum_i \beta_i x_i + e
\end{equation}
\par LASSO then provides the regression coefficients $\beta_i$ as a function of the regularization parameter $\lambda$, by minimizing $J(\beta_i)$:
\begin{equation}
J(\beta_i) = 1/N \sum_{j=1}^{N} (y_j- \sum_i \beta_i x_{i,j} )^2 + \lambda \sum_i \mid \beta_i \mid
\end{equation}

\par Increasing the penalty term $\lambda$ from zero (i.e. for LASSO switched off) to larger values eventually leads LASSO to set some coefficients to zero, which removes them from the selected model and thus reduces the dimension of the problem. The resulting regression coefficients $\beta_i$ can be compared to each other since variables $x_i$ are standardized at the start of the procedure. Our routine (based on the glmnet R package) uses a training set to learn from and a test set to search for the best $\lambda$, as well as a cross-validation procedure (10-folds here) to avoid overfitting. These are good practice rules that should be widely used in fitting procedures to increase robustness and generalization of the results, but are too rarely used in the literature.
\par Other regularization techniques exist, such as L2 regularization techniques as the Ridge regularization approach that is similar to LASSO but with a penalty term over the sum of the squared values of the regression coefficients. Ridge is known to be more efficient in the case of strongly correlated variables, but does not remove variables by setting some coefficients to null as LASSO. We will focus in the paper on the LASSO procedure and compare with ridge results.

\section{Results}
\label{Results}

\subsection{Bow shock}
\label{BS}

\par
\begin{figure}
    \centering
    \includegraphics[width=0.8\linewidth]{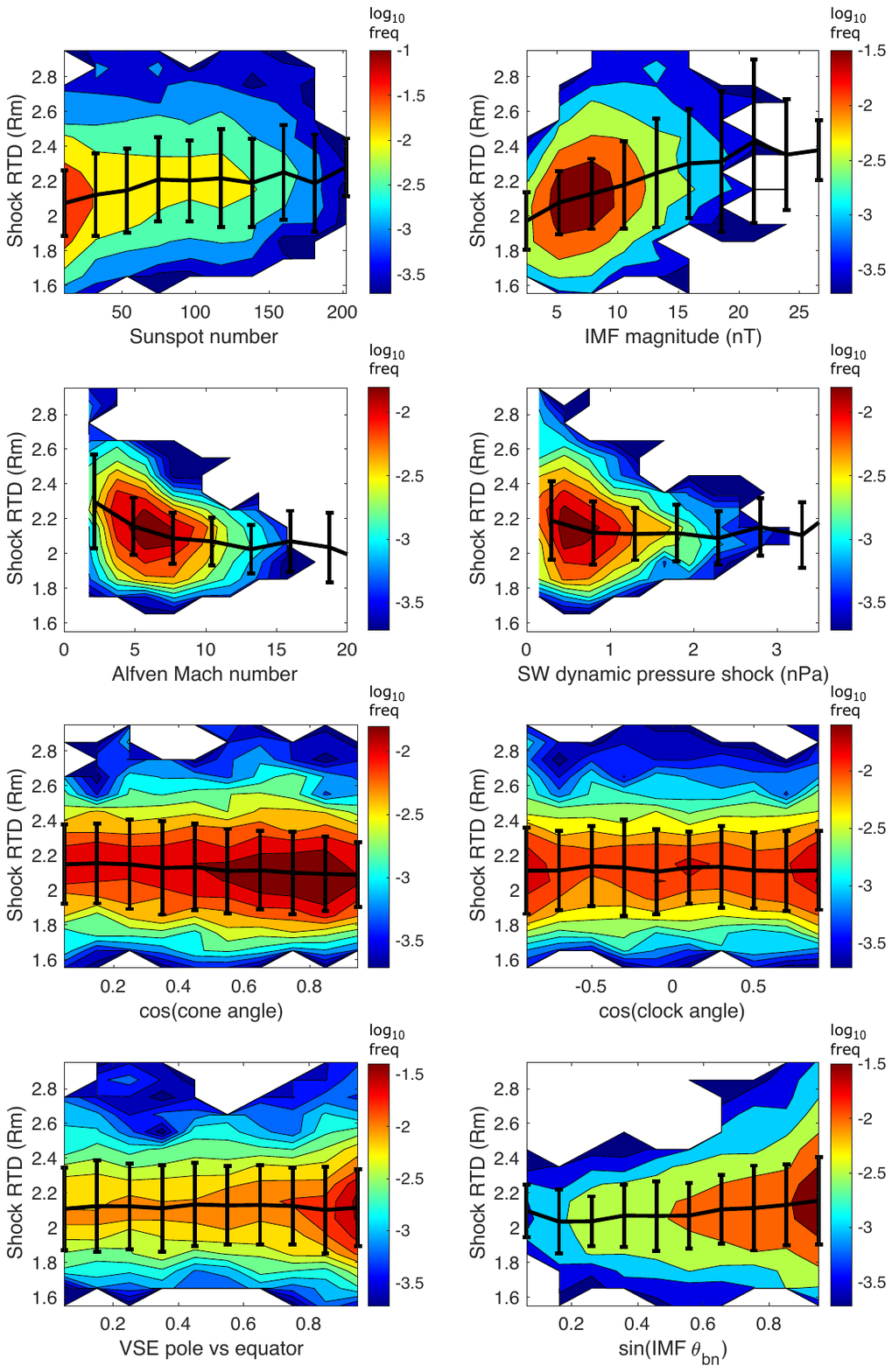}
    \caption{Logarithmic occurrence frequency of the extrapolated terminator distances of the Venus bow shock crossings as a function of several possible drivers: extreme ultraviolet fluxes (via sunspot number), IMF Alfven Mach number, IMF cone angle, IMF relative clock angle (see text), IMF magnitude, SW dynamic pressure, IMF clock angle, IMF $\theta_{bn}$ angle. Thick lines show the sliding median profiles with one standard deviation error.}
    \label{fig:shock_binette}
\end{figure}

 Figure \ref{fig:shock_binette} shows the logarithmic occurrence frequency of the extrapolated terminator distances of the Venus BS crossings as a function of several possible drivers. The drivers considered are: sunspot number (proxy for extreme ultraviolet fluxes), IMF Alfven Mach number ( $M_A=\frac{v_{sw}}{v_A}$ with $v_{sw}$ SW mean velocity and $v_{A}$ Alfv\'en speed), IMF cone angle (angle between the SW velocity vector and the IMF vector), IMF magnitude, SW dynamic pressure ($m_p n_{sw} v_{sw}^2$ with $m_p$ proton mass and $n_{sw}$ SW plasma density), IMF clock angle ($tan^{-1}(B_{Z_{IMF}}/B_{Y_{IMF}})$), IMF $\theta_{bn}$ angle between the local normal of the bow shock boundary at the crossing and the IMF vector, and IMF relative clock angle. This last parameter, called "pole vs equator", "relative clock angle" or simply "clock angle" of the crossings in older studies (e.g. \citet{alexander_cj:86} or \citet{russell_ct:88}) is here given by the absolute cosine of the angle between the IMF and the location of the crossing projected in the terminator plane (equivalent to the VSE frame representation) ; it can be used to show the influence of the mass loading of planetary ions picked up by the electric field, with possible enhanced BS location asymmetry in the direction perpendicular to the IMF. Note that we consider here cosine (for cone angle, clock angle, relative clock angle) or sine (for $\theta_{bn}$ angle) to use comparable parameter ranges and allow better insertion in statistical models, but the use of direct angles leads to unchanged qualitative results.

\par Overall one can observe that the Venusian BS $R_{TD}$ increases with:
\begin{itemize}
    \item increasing sunspot number: EUV fluxes are expected to increase atmospheric ionization rates and thus mass loading to the solar wind flow that interacts with a larger apparent obstacle ; EUV fluxes also heat the atmosphere and increase ionospheric scale heights and expand the atmosphere 
    \item decreasing Alfven Mach number: the BS is a fast magnetosonic standing wave created by the supersonic flow interacting with the ionized atmosphere of Venus, that propagates with a velocity determined by the magnetosonic Mach number ; magnetosonic Mach is a combination of Alfven Mach (that holds the Alfven velocity and SW velocity information) and of the sonic Mach (that adds the temperature information, not available here) ; 
    the Mach number both drives the jump conditions at the shock through Rankine-Hugoniot relations, and shapes the flaring of the BS with a shock cone inversely proportional to the Mach number from MHD modeling, thus modifying the cross section and the terminator distance of the BS (see e.g. \citet{michel_fc:65} or \citet{spreiter_jr:95})
    \item increasing IMF magnitude: the IMF intensity is expected to play a role through the Alfven Mach number with the Alfven velocity being proportional to the IMF magnitude, thus decreasing the magnetosonic Mach number when IMF increases; several authors suggested an importance of the IMF intensity or of its tangential component beyond its role through the Mach number, such as \citet{chai_l:14} or \citet{Wang_2024_a}, before removing it in later studies (\citet{chai_l:15} or \citet{Wang_2024_b}), while others keep it instead of Mach number \citep{Rollero_2025}
    \item decreasing solar wind dynamic pressure: a lower incident pressure compresses the whole induced magnetosphere less, letting the BS inflate further from the planet
    \item increasing cone angle (or decreasing cos(cone angle)): large cone angles are associated with increased $\overrightarrow{V} \times \overrightarrow{B}$ electric fields that lead to higher momentum transfer on heavy ions which then travel further and thus increase the size of the apparent obstacle to the solar wind
    \item increasing $\theta_{bn}$ angle (or its sine): quasi perpendicular BS crossings are expected to be located further from the planet since the wave velocity of fast mode magnetosonic waves is anisotropic and depends on the $\theta_{bn}$ angle \citep{khurana_kk:94}
\end{itemize}
\par On the other hand, the clock angle and the VSE pole vs equator (or "clock angle of the crossings") drivers show no clear trend according to figure \ref{fig:shock_binette}. These were however considered significant by several previous studies which analyzed the influence of the IMF orientation on the BS boundary asymmetries (see introduction), often explained in terms of mass loading asymmetric increase or geometrical effects related to the anisotropy of the magnetosonic waves velocity.

  \begin{figure}
      \centering
      \includegraphics[width=1\linewidth]{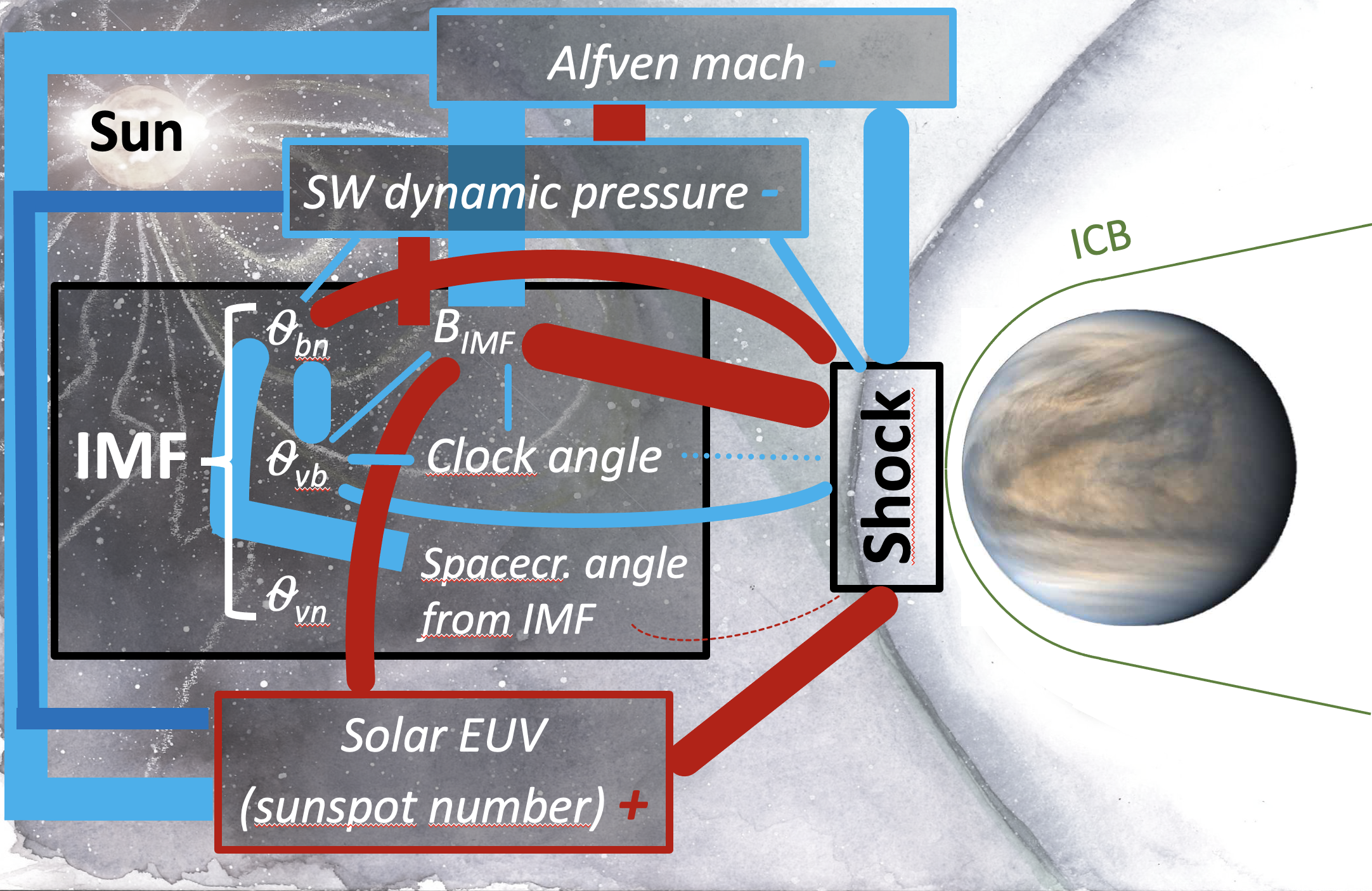}
      \caption{Schematic showing the complex inter-correlations of the possible drivers between themselves and with the Venus bow shock terminator distance. Blue and red lines correspond respectively to negative and positive Pearson linear correlation factors, while the thickness of the lines is proportional to the correlation factor. Dashed lines represent non-significant (p-value above 5$\%$) correlations. The background figure was adapted from a drawing by Anastasia Grigoryeva.}
      \label{fig:sketchshock}
  \end{figure}

\par Figure \ref{fig:sketchshock} shows a sketch of the Venusian interaction with solar wind plasma and fields, with a number of possible drivers of the BS included in this work. Blue/red lines correspond to negative/positive Pearson linear correlation factors, while their thickness is proportional to the correlation factor value. Dashed lines represent non-significant (p-values above $5\%$) correlations. The correlation factors are also given in Table \ref{table:correlationshock} that will be discussed later in the text. 
\par Figure \ref{fig:sketchshock} shows the same conclusions as above regarding the BS drivers apparent influence, since correlation factors essentially quantify the overall behaviors seen in scatter plots such as figure \ref{fig:shock_binette}. This representation and the previous figure suggest that both the Alfven Mach number and the IMF intensity are the most influent parameters, before EUV fluxes and then $\theta_{bn}$ angle or cone angle and SW dynamic pressure. However, this representation also shows how the possible drivers are cross correlated to each other with a number of apparent significant correlations, either expected from the parameters definition or not expected (e.g. Mach vs EUV fluxes, SW dynamic pressure vs EUV fluxes). These cross correlations may significantly alter conclusions regarding the real influence of minor drivers, or the ranking between the drivers by underestimating or overestimating their influence. For example, when EUV fluxes increase, Alfven Mach and SW dynamic pressure decrease at the same time, so that a part of the increase of the BS $R_{TD}$ induced by EUV increases may be due to other drivers that are modified in the same time, thus reducing the real influence of EUV fluxes. As shown by G22, when considering two variables of influence on a third parameter (here $R_{TD}$), their apparent influence given by scatter plots or correlation factors is overestimated (respectively underestimated) if the product between the three correlation factors (i.e. between variables and between variables and $R_{TD}$) is positive (respectively negative).

\par Table \ref{table:correlationshock} shows the direct / partial correlation factors of the possible drivers with respect to the BS $R_{TD}$ values. The comparison from direct to partial correlation factors on the same number of data reflects the impact of cross correlations discussed above. While $\theta_{bn}$ and SW dynamic pressure stay at a similar level of correlation with the BS distance, others show a significant decrease of correlation level with the BS $R_{TD}$ values when accounting for the cross correlations with the partial correlation approach: e.g. EUV and IMF magnitude partial correlations are reduced compared with initial direct correlation factors, due to cross correlations between EUV and IMF or Mach, so that considering their direct correlation factors only would actually lead to overestimating their influence on the BS location ; this is also the case for Alfven Mach and cone angle which in the end even show no significant partial correlation, due to a bias with IMF intensity. Moreover, relative clock angle, which was apparently not correlated from direct correlations, remains among the apparently influent drivers after taking into account cross correlations.

\par The results for the Alfven Mach number need to be discussed, since the same specific behavior appears for all methods. Unfortunately, the non availability of the ion temperature prevents us from including the magnetosonic Mach number, which is essential to describe the dynamics and shape of the Venusian BS, which is a fast magnetosonic standing wave propagating with a velocity determined by the magnetosonic Mach number. Here, the correlation factor of IMF intensity with the BS $R_{TD}$ is almost identical with the one between Alfven Mach number and $R_{TD}$, although slightly larger in absolute value ($0.38$ vs $0.37$). As a consequence, given their strong intercorrelation together (with a cross correlation factor of $-0.62$) due to the definition of Alfven Mach number, a "winner takes all" effect appears, with the strongest initially correlated parameter taking the whole after considering partial correlations. Thus IMF intensity remains significant while Alfven Mach appears non significant. In reality, one can expect that the magnetosonic Mach number, if included in the analysis, would be even better correlated with $R_{TD}$ than Alfven Mach number, which would thus remove IMF from the significant drivers after considering cross correlations. This is precisely what G22 showed in the context of the Martian BS, where IMF appeared non significant after removing the cross correlations with magnetosonic Mach number while it seemed significant at first, and this was also suggested by \citet{Wang_2024_b} at Venus. The above conclusions based on partial correlations are also true with the AIC or LASSO methods, where this "winner takes all" effect will be observed for Alfven Mach vs IMF magnitude.

\par Partial correlations thus suggest the following ranking of the Venus BS $R_{TD}$ drivers: 1) and 2) IMF intensity or (most probably) Mach number, as well as quasi perpendicular vs quasi parallel shocks through $\theta_{bn}$ (instead of EUV fluxes according to direct correlations) 3) EUV fluxes 4) SW dynamic pressure and relative clock angle at similar levels, while cone and clock angle are considered not significant.

\par If we consider an extreme case where all variables follow a power law instead of linear behavior, very similar results arise. $\theta_{bn}$, EUV and IMF remain among the most significant variables correlated with the BS distance, while the relative clock angle correlation appears again minor, clock angle stays non significant, and cone angle becomes barely significant close to the significance level. Note that Alfven Mach number however remains among the most significant correlations (largest partial correlation), and due in particular to the SW velocity information shared with SW dynamic pressure, SW dynamic pressure becomes non significant.

 \begin{table}
 \caption{Direct and partial correlation factor values between the possible drivers of the Venus bow shock boundary and the bow shock $R_{TD}$ distance.}
 \centering
\begin{tabular}{|c|c|c|}
\hline
Drivers & Direct correlation factor & Partial correlation factor\\
\hline
Alfven Mach & -0.37 & X$^{a}$ \\
EUV & 0.28  & 0.18 \\ 
IMF magnitude &  0.38 & 0.24 \\
sin(IMF $\theta_{bn})$  & 0.23 & 0.24 \\
 SW dyn. press  & -0.11 & -0.12 \\
 cos(cone angle)  & -0.12 & X  \\
 cos(clock angle)  & X & X \\
$\|$cos(Rel. clock angle)$\|$   & X & 0.10 \\
\hline
\multicolumn{3}{l}{$^{a}$X refers to a statistically non significant direct or partial correlation.}\\
\end{tabular}
\label{table:correlationshock}
 \end{table}

\par
\begin{figure}
       \centering
      \includegraphics[width=0.6\textwidth]{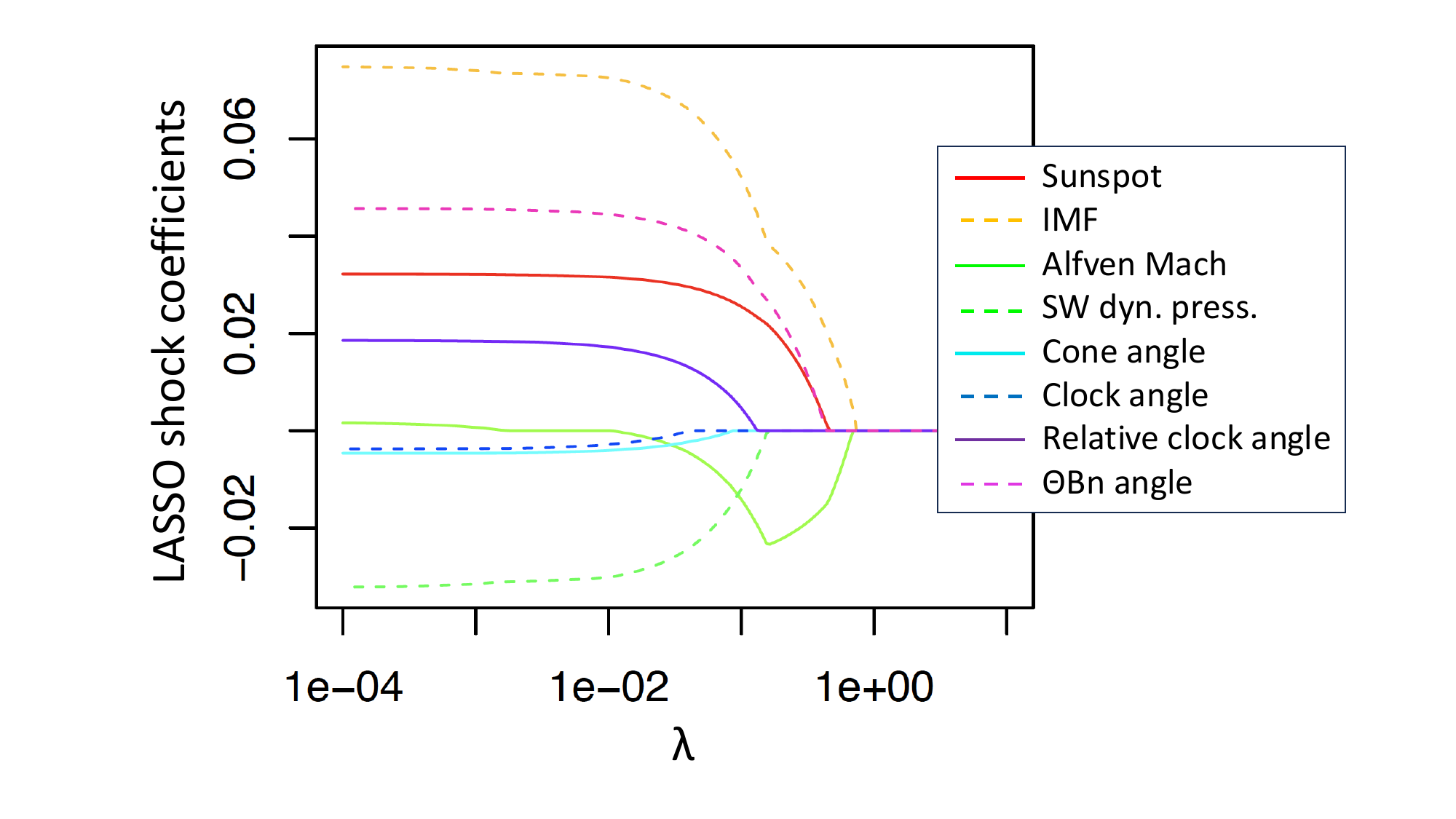}
      \caption{LASSO coefficients as a function of the penalty term $\lambda$ for the bow shock possible drivers.}
      \label{fig:LASSO_coeffvspenalty_shock}
  \end{figure}

 Results of the Least Absolute Shrinkage Selection Operator analysis are provided in figure \ref{fig:LASSO_coeffvspenalty_shock} and in table \ref{table:LASSOshock}, with the best standardized slopes of the multivariable linear regression (best final values in table and best values as a function of the penalty level in the figure) and a proposed reduction of the dimension. The slopes are provided after standardization of all drivers and of the BS $R_{TD}$ values ($X_{standardized}=\frac{X_{initial}-Mean(X)}{StandardDeviation(X)}$), so that all slopes are comparable and can allow us to suggest a ranking of the drivers.

   \begin{table}
 \caption{Standardized slopes provided by the LASSO approach for each of the Venus shock drivers considered. Ranks are added in parenthesis.}
 \centering
\begin{tabular}{|c|c|}
\hline
Drivers & Shock \\
\hline
Alfven Mach & 0.000 (X) \\
EUV & 0.032 (3) \\ 
IMF magnitude &  0.073 (1)  \\
sin(IMF $\theta_{bn}$)  & 0.045 (2)  \\
 SW dyn. press  & -0.031 (3)  \\
 cos(cone angle)  &-0.004 (6)  \\
 cos(clock angle)  & -0.003 (6) \\
$\|$cos(Rel. clock angle)$\|$  & 0.018 (5)  \\
\hline
\end{tabular}
\label{table:LASSOshock}
 \end{table}

\par Varying the penalty term from null to larger values leads to a variable set of slope values, with a growing number of variable slopes becoming null when increasing the penalty term. At first, Alfven Mach (due to the cross correlation with IMF) is already small and then disappears (before a increase, see below), before clock angle, cone angle and relative clock angle. The last removed parameters (the most robust drivers according to LASSO) are then, by chronological order and thus increasing importance, SW dynamic pressure, EUV fluxes, $\theta_{bn}$ and last IMF intensity. Let us note that removing IMF from the set of variables leads immediately to the Alfven Mach number becoming the primary and most robust driver of the BS $R_{TD}$ parameter. Moreover, as can be observed in the figure, when the penalty term is increased enough to reduce the SW dynamic pressure coefficient, then the Alfven Mach number coefficient rises again until a peak when the SW dynamic pressure coefficient becomes null, and then decreases again at the same time as the other drivers coefficients. This underlines the strong cross correlations (in particular via the SW velocity with SW dynamic pressure), which shows that several combinations of parameters can be used to reproduce the BS dynamics since several parameters share significant information. However, the best fit proposed by LASSO corresponds to a small $\lambda$ parameter in the early part of the figure where Alfven Mach number is considered less important than SW dynamic pressure.

\par The LASSO procedure then provides in the table a best set of variable slopes for a specific value of the penalty term (i.e. $8*10^{-4}$ for the BS), with the resulting slopes reflecting the relative importance of the drivers. Similar conclusions arise compared with the partial correlation analysis, except that more variables are kept by LASSO as significant (at the final level of penalty) even if considered less significant: relative clock angle, cone and clock angle.

\par Using ridge (L2 regularization technique) instead of LASSO leads to very similar results, with standardized slopes modified by maximum $10$ to $20\%$ but most importantly the ranking of the drivers is unchanged, except for the least influent parameters (clock angle and cone angle) whose coefficients ordering is reversed. Moreover, using power laws for the variables then linearized with a logarithm during the LASSO procedure also leads to an identical ranking of the drivers.

\par The third method used to analyze the relative importance of the BS drivers is the Akaike Information Criterion method described in section \ref{Methods}. Table \ref{table:AICshock} provides the results of this model selection approach applied to our dataset assuming at first linear combinations of the drivers. Removing the drivers one by one leads to models where part of the information is missing, the amount of information lost leading to the AIC score. The larger the score, the larger amount information lost if the corresponding variable is removed in the model.

   \begin{table}
 \caption{Akaike Information Criterion (AIC) score values for the ranking of the Venus shock drivers considered. Ranks are added in parenthesis.}
 \centering
\begin{tabular}{|c|c|}
\hline
Drivers & Shock \\
\hline
Constant & -4380 \\
Alfven Mach & X ((1) if IMF removed) \\
EUV & -4344 (3)  \\ 
IMF magnitude &  -4194 (1)  \\
IMF $\theta_{bn}$  & -4299 (2) \\
 SW dyn. press  & -4347 (4) \\
 Cone angle  & X \\
 Clock angle  & X \\
 Rel. clock angle  & -4369 (5) \\
\hline
\multicolumn{2}{l}{$^{a}$X refers to a non significant driver according to AIC.}
\end{tabular}
\label{table:AICshock}
 \end{table}

\par Consequently, the AIC scores suggest that IMF (or Mach if IMF is removed) is the most important driver of the BS $R_{TD}$ as its removal leads to the largest loss of information. Then, $\theta_{bn}$ also appears a strong driver, before EUV fluxes and SW dynamic pressure, while relative clock angle appears the last significant parameter whose score remains above the level that corresponds to the removal of a simple constant (i.e. $-4344$). Below this limit level one can find cone angle and clock angle that are non significant according to AIC. The use of BIC instead of AIC, which allows to increase the penalty term and further reduce overfitting leads to an identical ranking. The same is true for the use of power law behaviors for the variables instead of linear relationships.

\par To sum up, all three methods suggest the following similar conclusions, which points to the robustness of the results. The major driver of the BS location is most probably the 1) magnetosonic Mach number, but in the absence of reliable temperature measurements and due to significant cross correlations in our dataset (between IMF, Alfven Mach number, SW dynamic pressure and even EUV), the IMF intensity - or the Alfven Mach number that gets hidden just behind IMF - appears the most significant parameter of influence, with high IMF / low Mach numbers leading to increased BS $R_{TD}$. Then, the Venusian BS expands primarily for 2) quasi perpendicular shocks, 3) enhanced solar EUV fluxes and reduced SW dynamic pressures. A fifth possible driver is the relative clock angle influence with increased BS $R_{TD}$ for crossings occurring along the IMF direction rather than perpendicular to it, but the partial correlation approach considers it as non significant. Finally, LASSO is the only method to keep increased cone and clock angles of the IMF as significant drivers (at the best penalty level chosen) while partial correlations and AIC removed them.

\subsection{Ion Composition Boundary}
\label{ICB}

\par The same analysis were performed on the ICB crossings dataset that includes 916 crossings (only on the dayside).
\par
\begin{figure}
    \centering
    \includegraphics[width=0.8\linewidth]{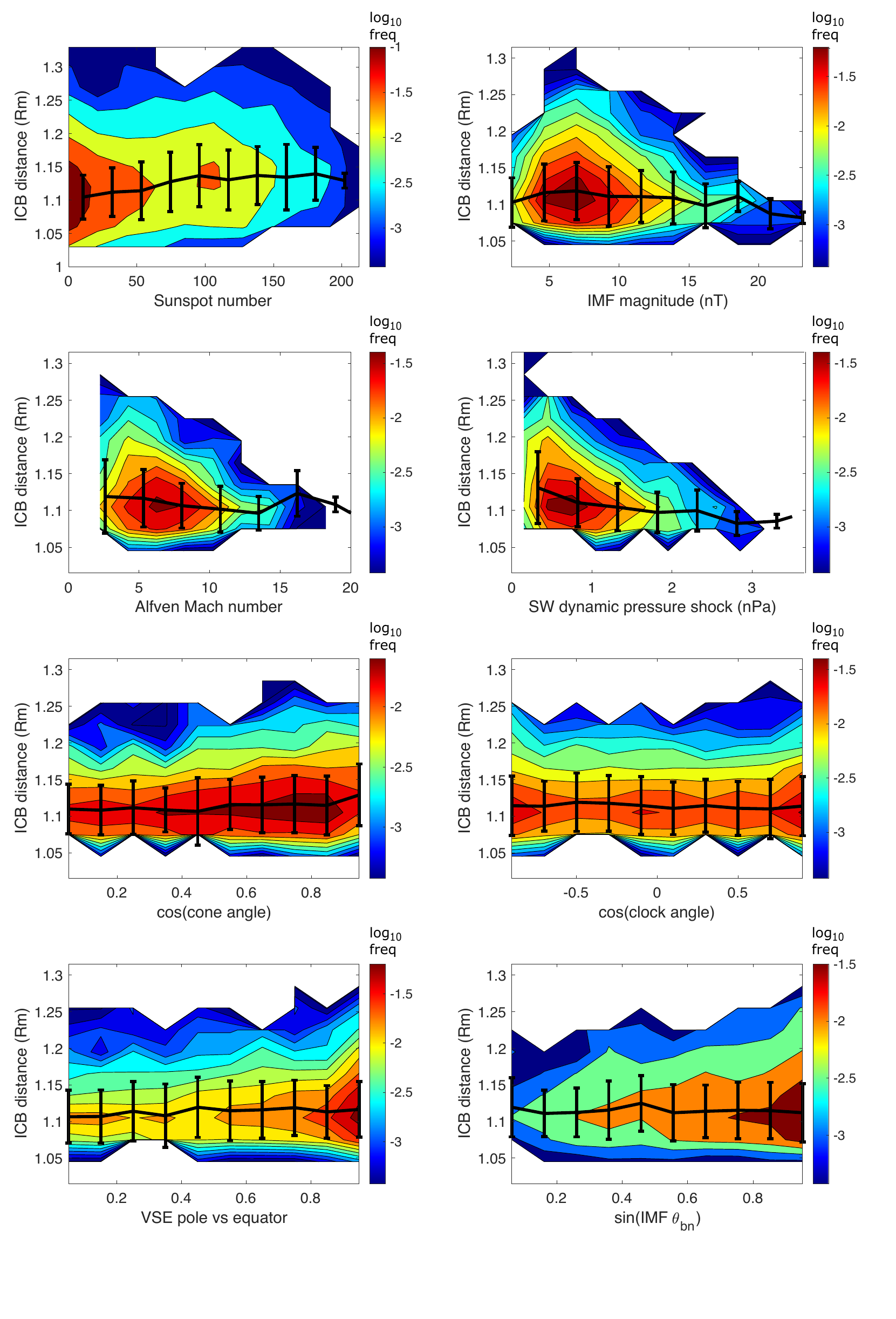}
    \caption{Same figure as fig. \ref{fig:shock_binette} for the dayside Ion Composition Boundary distance vs possible drivers.}
    \label{fig:scatterICB}
\end{figure}

 Figure \ref{fig:scatterICB} shows the logarithmic occurrence frequency of the ICB (assumed circular) distance as a function of the same possible drivers as those considered for the BS crossings in figure \ref{fig:shock_binette}.
\par The scatter plots suggest that the ICB expands with:
\begin{itemize}
    \item increasing sunspot number: EUV fluxes increase the ionization rates and heat the atmosphere, thus increasing ionospheric thermal pressure, which pushes the ionopause pressure balance vs external pressure components and can consequently push the ICB to higher altitudes due to the ionospheric expansion 
    \item decreasing solar wind dynamic pressure: a lower incident pressure compresses less the whole induced magnetosphere, thus expanding the Induced Magnetosphere Boundary, which lets the ionosphere expand and push the ICB further with a separation between the draped shocked magnetosheath plasma and the ionospheric plasma occurring at higher altitudes
    \item decreasing IMF magnitude, in particular at high values: similarly a reduced IMF intensity limits the compression of the induced magnetosphere, thus reducing the magnetic pressure in the magnetosheath which lets the ionosphere expand further
    \item decreasing Alfven Mach number (except at high Mach numbers with lower statistics): this behavior is due to the combined influences of IMF intensity and SW velocity, with the core of the distribution in the Mach number panel corresponding to the respective cores in the panels for IMF and SW dynamic pressure, where SW dynamic pressure has a clear influence (negative slope) while the ICB distance remains approximately constant with respect to IMF magnitude in the $5-10$ $nT$ range
    \item low cone angles and low relative clock angles: the influence seems smaller than the previous ones, but is in agreement with previous works (from Venus Express data analysis of MHD modeling, see \citet{xiao_zhang:18, Chang_2020, Xu_2022}) showing that low cone angle conditions weaken the pickup process and hence mass loading effects (due to reduced $\overrightarrow{V} \times \overrightarrow{B}$ convective electric field), which consequently weakens the upstream magnetic barrier, reduces the external pressure on the ionosphere that becomes less magnetized, and thus leads to higher ionopause altitudes \citep{xu_q:21}.
\end{itemize}
Moreover, the clock angle and $\theta_{bn}$ angle show no clear trends, as is expected for the $\theta_{bn}$ angle related to the shock geometry that should not impact ionospheric boundaries.

\par
\begin{figure}
      \centering
      \includegraphics[width=1\linewidth]{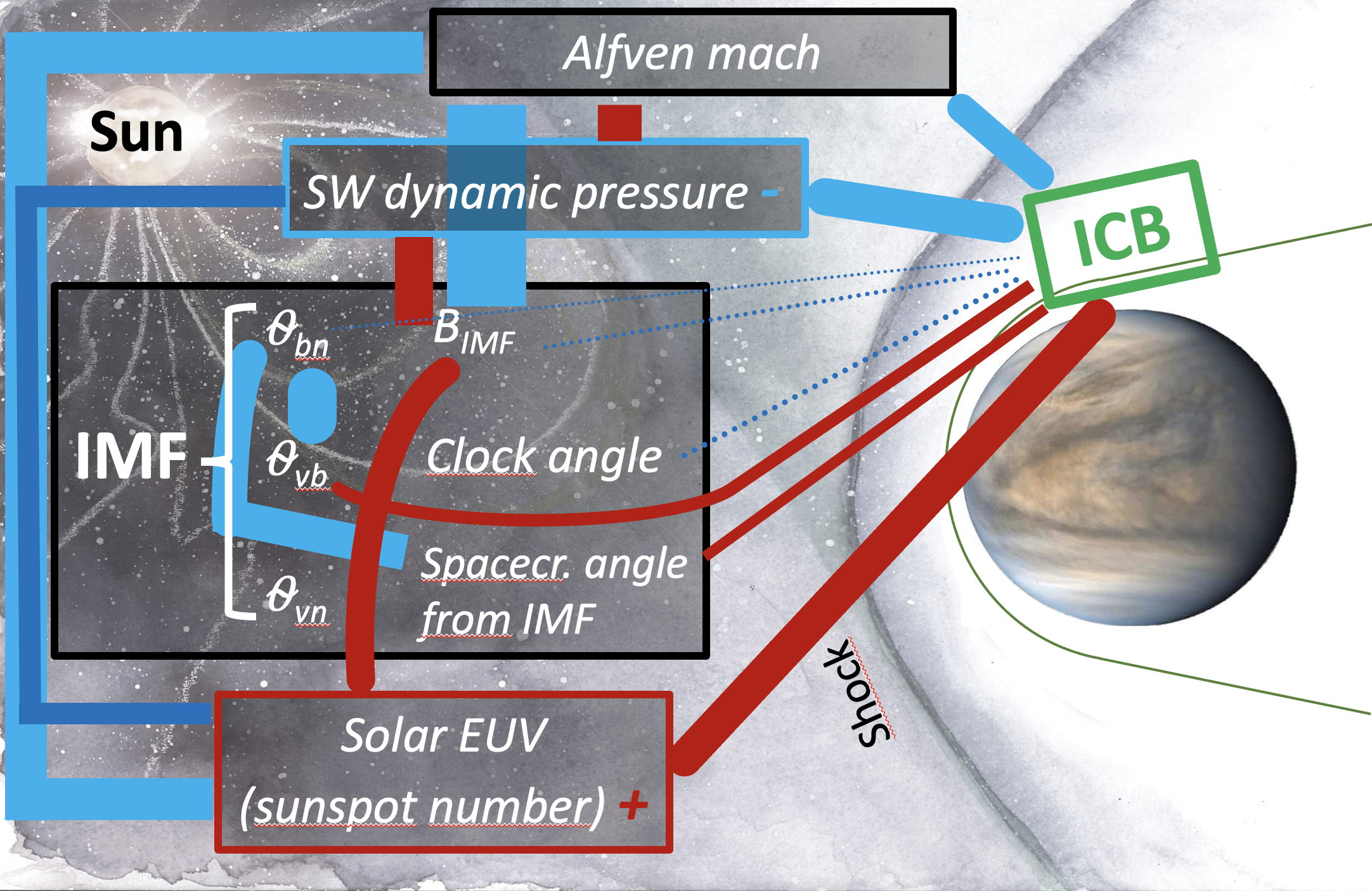}
      \caption{Same figure as fig. \ref{fig:sketchshock} for the possible ICB drivers.}
      \label{fig:sketchICB}
  \end{figure}

 Figure \ref{fig:sketchICB} represents graphically the direct and cross correlation factors of the possible drivers of the ICB with the same method as for the BS. One first remarks again the expected strong correlations between the Alfven Mach number and both the IMF and SW dynamic pressure (through the SW velocity). Similar cross correlations appear between the IMF parameters as in the BS dataset due to shared components in angle definitions, though with slight differences (e.g. no correlation observed for clock angle) due probably to the different (reduced in particular) dataset compared to the BS dataset. One can also see as for the BS dataset the significant cross correlations between EUV fluxes and SW parameters such as IMF intensity, SW dynamic pressure and Alfven Mach, which need to be considered to avoid underestimating or overestimating the quantitative and relative influence of these drivers on the ICB location.

\par The direct and partial linear correlation coefficients provided in table \ref{table:correlationICB} confirm the previous statements. The ICB distance is most significantly correlated with EUV, SW dynamic pressure, then Alfven Mach number to a lesser extent, and at a lower level with cone angle and relative clock angle. However, IMF appears non correlated at first with a more complex behavior in scatter plots.
\par Taking into account partial correlations between the drivers themselves and between the drivers and the ICB distance leads to a slightly different picture. Cone angle and relative clock angle keep small correlation levels, and clock angle as well as $\theta_{bn}$ keep non significant. On the other side, EUV becomes the most robust driver, and SW dynamic pressure appears less correlated due to shared information with the IMF intensity that finally remains among the significantly correlated parameters. Overall, EUV appears the most influent driver, while others are correlated to a lesser extent or not correlated ($\theta_{bn}$ and clock angle). Note that using power laws for all variables instead of linear relationships leads to similar conclusions: EUV remains strongly correlated, while all other variables become non significant except for relative clock angle (though not far from the significance limit) and SW dynamic pressure that remains significantly correlated with the ICB distance.

 \begin{table}
 \caption{Direct and partial correlation factor values between the possible drivers of the Venus ICB and the distance of the boundary.}
 \centering
\begin{tabular}{|c|c|c|}
\hline
Drivers & Direct correlation factor & Partial correlation factor \\
\hline
Alfven Mach &   -0.22 & -0.12 \\
EUV &  0.31 & 0.27\\ 
IMF magnitude &   X & -0.12 \\
sin(IMF $\theta_{bn})$  &  X & X\\
 SW dyn. press  &  -0.31 & -0.08\\
 cos(cone angle)  &  0.10 & 0.12 \\
 cos(clock angle)  &  X & X\\
$\|$cos(Rel. clock angle)$\|$   &  0.10 & 0.13 \\
\hline
\multicolumn{3}{l}{$^{a}$X refers to a statistically non significant direct or partial correlation.}\\
\end{tabular}
\label{table:correlationICB}
 \end{table}

\par
\begin{figure}
      \centering
       \centering
      \includegraphics[width=0.6\textwidth]{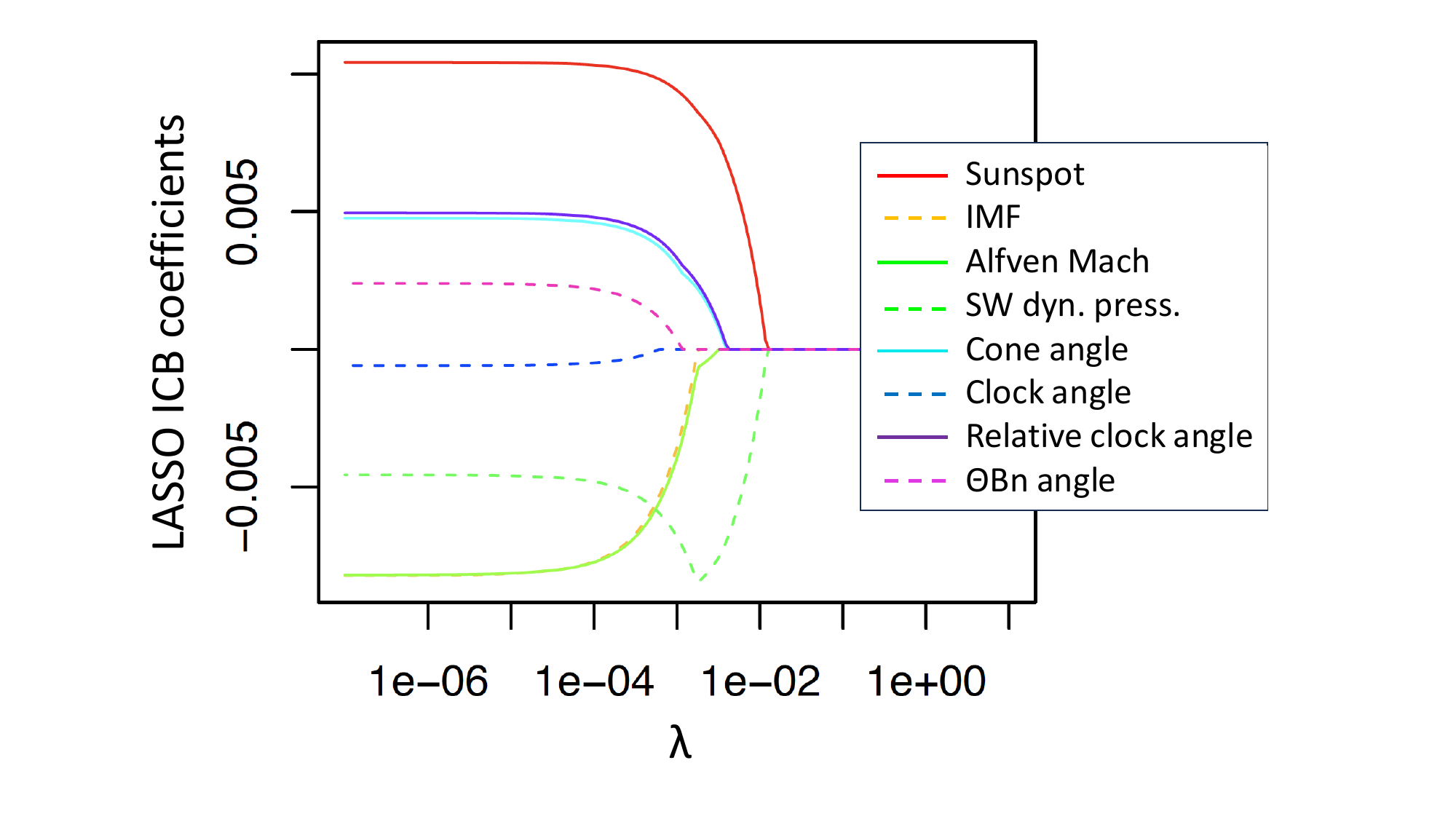}
      \caption{Same figure as fig. \ref{fig:LASSO_coeffvspenalty_shock} with ICB drivers LASSO coefficients as a function of the penalty term.}
      \label{fig:LASSO_coeffvspenalty_ICB}
  \end{figure}

 The LASSO analysis results are provided in figure \ref{fig:LASSO_coeffvspenalty_ICB} and table \ref{table:LASSOICB} with the standardized slopes of a multivariate linear (at first order) regression of the ICB distance with respect to the drivers considered. Increasing the penalty term in the LASSO procedure leads to the following behavior for the standardized slopes of the possible drivers. First the clock angle coefficient is very small from the start and decreases further just before the $\theta_{bn}$ coefficient, both then staying at low levels. Then the cone angle and relative clock angle share a similar evolution with similar levels of their respective coefficients that decrease together, while Alfven Mach number and IMF intensity coefficients start from higher values but decrease rapidly. 
\par The SW dynamic pressure shows a more complex evolution, with an initial coefficient at similar levels to cone and relative clock angles, but its coefficient increases at first until a peak when the previous group of variable coefficients (cone and relative clock, as well as Alfven Mach and IMF intensity) decreases, before it decreases for stronger penalty levels in the same time as the sunspot number that remains the strongest driver. The influence of SW dynamic pressure is thus reduced as long as other variables such as Alfven Mach number and IMF (that are strongly correlated to it) are also present in the statistical model, so that its relative influence depends on the presence or not of these variables that share significant information with it. This is the reason why considering only scatter plots or direct correlation factors (that do not take into account cross correlations) suggests that SW dynamic pressure is a primary driver (as considered by S23).
\par The best penalty level considered by LASSO is small (i.e. $9.4*10^{-5}$) which corresponds to a state where the final ranking of the coefficients (given in the table) is the following, with none of the parameters removed: 1) Sunspot number 2) Alfven Mach number or IMF intensity 3) SW dynamic pressure, cone and relative clock angles 4) $\theta_{bn})$ and 5) clock angle. Note that using a L2 regularization with the Ridge approach leads to almost unchanged results for the best coefficients (by about $5\%$). Using power laws instead of linear behaviors leads to the same ranking for the drivers, with the Alfven Mach number and IMF intensity LASSO coefficients with respect to the ICB distance however closer to the EUV coefficient level.

   \begin{table}
 \caption{Standardized slopes provided by the LASSO approach for each of the Venus ICB drivers considered. Ranks are added in parenthesis.}
 \centering
\begin{tabular}{|c|c|}
\hline
Drivers & ICB \\
\hline
Alfven Mach  &  -0.008 (2) \\
EUV &  0.010 (1)\\ 
IMF magnitude &   -0.008 (2) \\
sin(IMF $\theta_{bn}$)  &  0.002 (7) \\
 SW dyn. press  &  -0.005 (4) \\
 cos(cone angle)  & 0.005 (4) \\
 cos(clock angle)  &  -0.0005 (8)\\
$\|$cos(Rel. clock angle)$\|$  &  0.005 (4) \\
\hline
\end{tabular}
\label{table:LASSOICB}
 \end{table}

\par The Akaike Information Criterion results shown in table \ref{table:AICICB} provide the following conclusions regarding the relative importance of the ICB distance drivers at Venus. The AIC scores tell us that the primary driver is by far solar EUV fluxes. Then, one can find close to each other relative clock angle, cone angle, IMF intensity and Alfven Mach number. Finally, one can find SW dynamic pressure and at last  the IMF $\theta_{bn}$ that is close to the significance level (level below which removing a variable leads to less loss of information than a simple constant). Clock angle is then considered not significant by AIC. 

\par Note that removing IMF intensity from the procedure leads to SW dynamic pressure becoming the second most important driver behind EUV fluxes, while the Alfven Mach number becomes not significant. As mentioned above, much information is thus shared between these SW variables: SW dynamic pressure, IMF intensity and Alfven Mach number. Using power laws for the variables leads to an identical ranking as the linear behavior.

\par The combined results from the three methods used here suggest the following conclusions regarding the dayside Ion Composition Boundary drivers. First, the main driver of the ICB location is clearly the EUV fluxes influence through the ionization rates and thermal pressure increase. Then, a number of variables also influence to a lesser extent the ICB location: the solar wind draping and induced compression of the Venus ionized atmosphere has some impact on the ICB distance on the dayside through several parameters that are strongly cross correlated to each other (IMF, Alfven Mach number and SW dynamic pressure), while IMF induced asymmetries remain statistically significant (pole vs equator asymmetry via relative clock angle, and conventional electric field $- \overrightarrow{V} \times \overrightarrow{B}$ induced asymmetries via the cone angle influence).


    \begin{table}
 \caption{Akaike Information Criterion (AIC) score values for the ranking of the Venus ICB drivers considered. Ranks are added in parenthesis.}
 \centering
\begin{tabular}{|c|c|}
\hline
Drivers &  ICB \\
\hline
Constant & -6128 \\
Alfven Mach &   -6117 (5) (X if IMF removed) \\
EUV &  -6061 (1) \\ 
IMF magnitude &   -6116 (4) \\
IMF $\theta_{bn}$  &  -6126 (7) \\
 SW dyn. press  &  -6123 (6) (2 if IMF removed)\\
 Cone angle  &  -6116 (3)\\
 Clock angle  &  X\\
 Rel. clock angle  &  -6113 (2)\\
\hline
\multicolumn{2}{l}{$^{a}$X refers to a non significant driver according to AIC.}
\end{tabular}
\label{table:AICICB}
 \end{table}

\section{Discussion}
\label{Discussion}

\subsection{Comparison with previous studies}
 
\par Compared with previous studies of the BS location, our results suggest that at least Mach number, $\theta_{bn}$ but also EUV and SW dynamic pressure should be included in parametric models (e.g. \citet{Wang_2024_b} did not include $\theta_{bn}$, \citet{Rollero_2025} did not include EUV). SW dynamic pressure appears as a significant but second order driver just behind EUV fluxes, due to strong cross correlations with other solar parameters. S23 considered SW dynamic pressure a major driver of the BS together with solar cycle, but did not consider quasi-perpendicular vs quasi-parallel shocks as a significant driver. 

\par Regarding the ICB location, EUV and SW dynamic pressure were considered by previous authors as the major drivers (\citet{alexander_cj:86} ; \citet{angsmann_a:11}, \citet{wei_y:12}). However, EUV appears in our conclusions as the primary and most robust driver, with also some influence of SW parameters that are either IMF intensity and Alfven Mach together or SW dynamic pressure alone (if IMF intensity is not included), as well as IMF induced asymmetries (pole vs equator asymmetry via the relative clock angle) or low cone angles that reduce the magnetic pile-up \citep{xiao_zhang:18} and thus allow the ionopause to expand as observed by \citet{xu_q:21}. This underlines the need to consider a large number of drivers together to identify their real importance due to significant cross correlations between them.
\par Note that replacing the SW dynamic pressure by the SW proton fluxes as proposed by \citet{Rollero_2025} leads to unchanged drivers ranking for the BS (and unchanged quality of the best AIC or LASSO models), while proton flux becomes the third driver for the ICB behind sunspot and relative clock angle (with a slightly reduced quality of the resulting AIC best model). Our results are thus consistent with \citet{Rollero_2025} on the possible use of protons flux as one of the boundary drivers.

\par Based on the same dataset, S23 did not consider EUV as a major direct driver of the ICB distance, but suggested different behaviors and drivers during solar minimum vs maximum periods. Actually, separating the dataset into solar minimum and maximum datasets hides the importance of the EUV fluxes. Using our methods on the same dataset but separated into two periods indeed pushes EUV fluxes much further in the drivers rankings for both the BS and ICB (or even removes it from the significant drivers), since the EUV fluxes influence is mostly related to a seasonal influence that takes place on large periods, that can be hidden behind more dynamic drivers (i.e. SW related parameters) if shorter periods of time are considered in the analysis. Using two different values of the BS conic eccentricity for solar minimum vs maximum conditions as performed by S23 (also used in this study) does however not reduce the apparent influence of EUV fluxes: using a single eccentricity as done by \citet{Rollero_2025} (with a mean value of $1.047$) leads to the same ranking for the BS drivers, except that SW dynamic pressure gets ranked just above sunspot number instead of the reverse situation with two eccentricities, but both drivers stay at similar levels of AIC scores or LASSO coefficients. 

\par Our conclusions underline the importance of considering a large set of possible variables together to investigate which ones carry the most information on the final parameter studied, here the distance / extrapolated terminator distance of plasma boundaries. The expected or unexpected presence of cross correlations needs to be taken care of with precaution, which can be strongly ameliorated with the methods proposed in this paper. When considering the influence of minor drivers, cross correlations can lead to apparently significant drivers from scatter plots or direct correlation factors to non significant in the end (e.g. cone angle for the BS) or vice versa leading to possible significant drivers that did not appear as significant at first (e.g. relative clock angle for the BS or IMF intensity for the ICB).

\par One can mention the debate regarding the true influence of the SW dynamic pressure, in particular on the shock and during solar minimum conditions at Venus. First, several studies found no clear influence of the SW dynamic pressure at all on the Venusian shock (e.g. \citet{chai_l:14}, \citet{chai_l:15}, \citet{zhang_tl:04}, \citet{Wang_2024_a}, \citet{Wang_2024_a}) while others did. Second, some studies also mentioned a differential impact depending on solar conditions. \citet{martinez_c:08} found that the Venusian shock position is insensitive to changes in the dynamic pressure of the SW, at least during solar minimum. However, one can note limitations: the dataset was more limited (248 bow shock events only) and all occurred during solar minimum conditions. Later, S23 found in their dataset a minimal effect of SW dynamic pressure on the Venusian bow shock during solar minimum conditions compared to solar maximum conditions. They attributed this differential influence (seen also in the IMF influence on the shock) to a change of magnetization state of the ionosphere during solar minimum conditions. Indeed, it has long been known that the structure of the induced magnetic field in Venus ionosphere enables it to be classified as either magnetized or unmagnetized, as investigated recently in details by \citet{byrd_s:24}. The ionospheric thermal pressure at Venus is often comparable to the SW dynamic pressure, so that the induced fields are weaker than required to balance the SW by themselves, while these are stronger than required to achieve pressure balance at Mars. If the ionosphere is in a magnetized state, which happens more often in solar minimum conditions with low ionospheric thermal pressure, the total ionospheric pressure may include a part of the IMF induced external pressure, which then could help maintaining the boundaries altitude during increased SW dynamic pressure events. As a consequence, the true influence of SW dynamic pressure on Venus plasma boundaries could be reduced in solar minimum conditions in Venus ionosphere (probably not at Mars).
\par In our study, that revisits the work by S23, our conclusions are different based on methods that better consider cross correlations between multiple parameters of influence. The LASSO approach might indicate a smaller influence (with smaller normalized slopes) of SW dynamic pressure during solar minimum conditions with respect to other drivers of the shock. However, contradictory results arise when comparing with the AIC approach, that suggests no clear influence of solar conditions on the SW dynamic pressure ranking. No clear trend appears regarding the SW dynamic pressure influence on the ICB location during solar minimum vs maximum conditions. Moreover, separating the dataset into minimum and maximum solar conditions datasets reduces the statistics and hides at least partially the effects of drivers whose influence occurs on large periods (such as the seasonal influence of EUV), which also concerns the SW dynamic pressure that also shows long period variations. This is clearly seen with the EUV fluxes whose influence appears strongly reduced (and gets eventually invisible) as mentioned above. Thus, one should take strong precautions when analyzing reduced datasets such as solar minimum vs maximum conditions, since several drivers influence get easily hidden or poorly estimated, which applies not only to our methods but obviously also to classic direct analysis performed in most studies. The differential influence of SW dynamic pressure on the Venusian boundaries thus remains to be further studied, even though one expects that the magnetization state of the ionosphere could impact their dynamics during solar minimum conditions.

\par Our results also underline the difficulty to identify clearly the influence of asymmetries due to the IMF orientation, since most SW parameters are cross correlated to each other and in particular with $\theta_{bn}$, IMF intensity or Alfven Mach number. In this case, comparing several methods appears important since individual methods often provide a specific point of view with slightly different conclusions from a method to another, so that keeping only the common and thus most robust conclusions helps to increase the confidence in the results. Previous authors investigated extensively these aspects at Venus during the Pioneer era, but did not consider cross correlations with other parameters and used datasets with a limited number of crossings as well as a unique approach \citep{alexander_cj:86,russell_ct:88}. Here, our conclusions confirm a probable influence (beyond the strong $\theta_{bn}$ influence) of relative clock angle (i.e. pole vs equator asymmetry in the IMF projected frame) for the BS and the ICB as well as of cone angle for the ICB, while an influence of cone and clock angles on the BS is not excluded. The small statistical significance of cone angle influence on the BS suggests that the convective electric field may have less influence than expected in previous works. \citet{xu_q:2023} observed a small solar wind control (compared to the EUV influence) of the pickup ion plume in the dayside magnetosheath of Venus, suggesting a moderate solar wind control of the planetary mass loading to the solar flow compared to EUV induced ionization, which is in agreement with the reduced impact of cone angle (and thus of the convective electric field) observed in our analysis.

\par One can also note that, even though our analysis is not meant to provide functional fits for the BS and ICB locations with respect to their physical drivers, the AIC and LASSO methods provide simple linear (or power law) fits which appear efficient. One can for example compare with the model functions by \citet{Rollero_2025}, either by comparing directly the data-model RMS distance for the ICB, or by reverse engineering from our $R_{TD}$ value for the BS and compare the data-model RMS distance (assuming the same $X$ value given by the real crossings). Model to data RMS distance values of $\sim 0.26$ $R_V$ (vs $0.32$ $R_V$ for \citet{Rollero_2025}) and $\sim 0.034-0.035$ $R_V$ (vs $0.039$ $R_V$ for \citet{Rollero_2025}) are thus obtained respectively for the BS and ICB boundaries from the AIC or LASSO fit functions (see tables \ref{table:LASSOshock} and \ref{table:LASSOICB} for LASSO coefficients). This does not mean our simple linear fits should be used instead of other fit functions from the literature, since 1) more variables are considered in our AIC and LASSO procedures (\citet{Rollero_2025} included fewer variables in their final models) which leads to lower RMS 2) one may want to use functional forms that reflect the physical processes at work whose dynamics (linear, power law like or other behavior) and timescales (seasonal effects, SW induced dynamics...) may be different. However, this confirms at least our assumption that a simple linear (or power law linearized) approach allows to catch a significant part of the boundary dynamics and helps understanding the influence of the various drivers.

\subsection{Comparison between Mars and Venus bow shock drivers}

\par We compare in table \ref{table:ChocMarsVenus} the rankings of the Venusian and Martian BS drivers as suggested from the three following methods (based on this study and G22): partial correlations, Akaike Information Criterion, LASSO. As expected, both planetary boundaries share a number of common characteristics, due to the similar nature of both planetary interactions, i.e. bodies with an EUV (mostly for the dayside) ionized atmosphere without an intrinsic global magnetic field interacting with a supersonic and superalfvenic solar wind plasma. The major BS drivers at both planets are overall (with no common ranking however): increased solar EUV fluxes, most probably reduced magnetosonic Mach number (or IMF at Venus, that is strongly correlated to Mach number, since our dataset did not allow to include magnetosonic Mach number) and quasi perpendicular shocks versus quasi-parallel shocks (via $\theta_{bn}$ angle). Then, SW dynamic pressure keeps also a significant second order independent driver of the BS location (at similar levels as crustal fields at Mars, that do not exist at Venus).

\par Differences arise however when comparing the rankings of the BS drivers for both planets. First, solar EUV illumination has more influence at Mars where it appears as the major driver while it is ranked third by all methods at Venus behind 1) IMF or Mach and 2) $\theta_{bn}$. This difference is in agreement with the stronger seasonal effects widely discussed in the literature existing at Mars than at Venus, whose eccentricity and rotation axis tilt angle are almost null compared to Mars (eccentricity of $0.1$ and $25^{\circ}$ inclination). \citet{peter_k:25} reported an irradiance variability during the solar cycle number 24 (from 2009 to 2020) of a factor of almost $3$ for Mars (with irradiance levels from $1.5$ to $4*10^{14}$ $s$$^{-1}$$m$$^{-2}$) and less than 2 for Venus (with irradiance levels from $9$ to $15*10^{14}$ $s$$^{-1}$$m$$^{-2}$). Comparing the BS normalized regression slopes provided by the LASSO approach from G22 and from our study (normalized again by the planet radius) also suggests that, beyond the stronger variability of EUV fluxes at Mars, the relative influence of EUV on the BS $R_{TD}$ dynamics at Mars (i.e. the efficiency of EUV fluxes to modify the BS location) is stronger with a slope three times larger.
\par Then, the relative clock angle (or pole vs equator asymmetry) seems slightly more important at Venus than at Mars, while the cone angle influence also appears stronger according to LASSO (with a factor of 5 of difference between the normalized slopes). This might be related to an enhanced convective electric field at Venus vs Mars discussed by \citet{curry_s:15} that would induce larger pickup ion fluxes and would be maximized by larger cone angles. However, how an enhanced convective electric field would increase the influence of relative clock angle or cone angle at Venus remains to be studied. Moreover, the cone angle influence needs confirmation since it does not appear as significant according to AIC nor partial correlation due to biases with other more influent drivers. Meanwhile, clock angle appears influent at Mars and not at Venus, but G22 commented the influence of clock angle saying that precautions should be taken due to both highly inhomogeneous distributions and reduced influence after taking into account cross correlations. Overall, except for the strongly influent $\theta_{bn}$ angle, and for the relative clock angle that appears as a minor but robust possible driver based on several methods, the clear influence of other IMF related angles such as cone and clock angle on the BS remains to be analyzed further.

 \begin{table}
 \caption{Comparison between the Venus and Mars bow shock driver rankings according to three different methods: partial correlation values, Akaike Information Criterion, LASSO.}
 \centering
\begin{tabular}{|c|c|c|c|c|c|c|}
\cline{2-7}
\multicolumn{1}{c|}{} & \multicolumn{3}{c|}{Venus} & \multicolumn{3}{c|}{Mars}\\
\hline
Shock drivers & Partial correlation & AIC score & LASSO & Partial correlation & AIC score & LASSO \\
\hline
Mach & X$^{a}$ or 1 & X or 1 & X or 1 & 3 & 3 & 2 \\
EUV & 3& 3& 3& 1& 1 & 1\\ 
IMF magnitude & 1 & 1 & 1 & 6 & 9 & 6 \\
IMF $\theta_{bn}$  & 1& 2& 2& 2& 2& 3\\
 SW dyn. press  & 4& 4& 3& 4& 4 or 5& 4\\
 Cone angle  & X& X & 6& X& X& 8\\
 Clock angle  & $X$& X& 7& 6& 8& 7\\
 Rel. clock angle  & 4 or 5& 5& 5& 6& 7& 7\\
 Crustal fields$^{b}$  & & & & 5 & 4 or 5& 5\\
\hline
\multicolumn{7}{l}{$^{a}$X refers to a non significant driver according to the method.} \\
\multicolumn{7}{l}{$^{b}$Crustal fields were only considered for the Mars case.}
\end{tabular}
\label{table:ChocMarsVenus}
 \end{table}

\subsection{Ion composition boundary vs bow shock behavior}

\par Analyzing the combined behavior of plasma boundaries beyond individual behaviors is also important to get a better understanding of the global dynamics of the planetary magnetospheres. 

\par First, comparing the drivers of both boundaries leads to the following remarks. Solar EUV and SW characteristics significantly influence the BS and ICB boundaries, but EUV is the only common strong driver for both boundaries. The most influent SW parameters are different for each, with IMF (or Mach) and $\theta_{bn}$ for the BS, while a number of SW parameters influence the ICB to a lesser extent than EUV fluxes (except $\theta_{bn}$ that is close to the non significance level). Comparing the LASSO slope values of the most common SW drivers reveal for example a stronger apparent influence of the IMF (or Mach ; $+370\%$), SW dynamic pressure ($+220\%$) and EUV fluxes ($+50\%$) for the BS with respect to the ICB, even after normalizing for the mean distance of both boundaries. This confirms the importance for including these drivers to understand the BS dynamics, as well as the overall larger variability of the BS boundary location.
\par Second, the distributions of the BS and ICB locations (BS $R_{TD}$ or ICB distance) reveal several interesting features:
\begin{itemize}
    \item both boundaries are characterized by a strong peak and strong tails compared with classical gaussian distributions, given their respective kurtosis ($9.0$ and $6.5$ respectively for the BS and ICB, indicating strongly leptokurtic distributions). Both boundaries thus have an overall stable location, with however a number of extreme events leading to a strong tail
    \item both boundary distributions are highly asymmetric and skewed towards larger distances rather than smaller distances, with large positive skewness values of respectively $1.3$ and $1.4$ for the BS and ICB ; this means both boundaries have a tendency for expansion rather than compression
    \item the dispersion of the BS location is much larger than the ICB (by a factor 3 or 5 if absolute or standardized variations are considered), confirming the BS is more compressible or expandable than the ICB, as previously mentioned by \citet{Rollero_2025}
\end{itemize}
\par Third, the interdistance between both boundaries is also indicative of the differential or common behavior or both boundaries. We calculated the standardized BS $R_{TD}$ and ICB distance and then derived the "interdistance" (thus focused on the interdistance extrapolated at the terminator) from these standardized locations. Note that keeping the absolute interdistance only leads to the same conclusions as analyzing the BS location due to its larger variability. The third (i.e. skewness) and fourth (i.e. kurtosis) moments of the resulting standardized interdistance distribution show that the relative distance between both boundaries is also strongly peaked with significant tails induced by extreme events, and less skewed towards expansion than the individual distributions. Removing extreme events (beyond $3$$\sigma$) even removes the skewness of the interdistance distribution. This suggests that both boundaries usually share a common behavior, since they are mostly driven by the same drivers, but extreme boundary excursions may be induced by different conditions.

\par We applied the same approach for the interdistance (standardized or not) as for the individual boundaries' dynamics by applying both the LASSO and AIC algorithms. Both methods suggest the same ranking for the main drivers of the BS to ICB interdistance dynamics: 1) IMF intensity (or Mach if IMF is removed) that impacts both boundaries 2) $\theta_{bn}$ that strongly influences the BS 3) sunspot number that also influences both boundaries. Then, the ordering of the other influent parameters depends on the method and the type of interdistance chosen (standardized or not), with overall SW dynamic pressure and relative clock being the most secondary order significant drivers. The drivers ranking of the interdistance thus remains closely related to the BS drivers ranking, even when normalizing to the average boundary location. This shows that the Venusian magnetosheath size is maximum for strong IMF intensities, quasi-perpendicular shock conditions (thus leading to sheath asymmetries) and strong incident EUV fluxes, while other parameters can also alter the sheath size to some extent. One can note that the mean absolute interdistance between both boundaries (at the terminator plane given the use of $R_{TD}$ for the BS) is of the order of $70$ ion inertial lengths (with $25-75$ percentiles from $55$ to $85$ $c/\omega_{pi}$), thus confirming the large size of the induced magnetosphere of Venus.

\subsection{The extreme boundary locations}

\par The statistical analysis performed in this study are based on the full dataset of boundary crossings available to us. One question we can ask from this dataset is: Which are the conditions leading to extreme excursions of the boundaries, and what is the impact of the extreme events on the interpretation of the overall dataset ?
\par We selected the extreme events based on the standardized BS $R_{TD}$ and ICB distances, by setting a threshold of $3\sigma$. This corresponds to respectively $68$ and $36$ BS and ICB crossings. Among them, only one corresponds to an extreme close crossing of the shock, all other extreme events correspond to far shock crossings or far ICB crossings, which is coherent with the tendency of the boundaries to expand rather than to be compressed. One can also note that none of the BS and ICB extreme events occurred at the same orbit, thus suggesting that extreme events are induced by different conditions despite an overall common behavior given the conclusions above.
\par Analyzing the distributions of the various possible drivers during the extreme boundary crossing events reveals that: 1) the IMF and Mach number conditions were on average severe during the extreme BS crossings (above $1\sigma$ deviation level) and SW dynamic pressure as well as EUV fluxes showed significant deviations from their mean behavior ($>0.5\sigma$ level) 2) a number of drivers showed moderately strong/low conditions ($>0.5\sigma$ deviation) for the ICB extreme events such as SW dynamic pressure, Mach number, EUV conditions, while IMF conditions where close to normal ($<0.5\sigma$ deviation).
\par In addition, it appears that most of the BS extreme events correspond to conditions where several parameters showed severe conditions rather than extreme conditions for a unique driver, with $67\%$ cases with at least $2$ parameters above $1\sigma$ deviation levels at the same time (and half with at least $3$ severe concomitant conditions). One can note however that severe IMF conditions were observed in many cases, combined with other more or less severe drivers conditions, which suggests IMF plays a significant role in the appearance of extreme BS excursions. On the other side, the ICB extreme events often (half of the events) occurred when a unique parameter (essentially sunspot number) showed severe conditions above $1\sigma$ deviation, or otherwise when several parameters showed more moderate deviations. Extreme ultraviolet fluxes thus appear as the main source for extreme ICB expansions.
\par Finally, we investigated the impact of extreme excursions of the boundaries (beyond $3\sigma$ deviation) on the overall ranking of the drivers by comparing AIC and LASSO results with or without including these events (i.e. $68$ BS and $36$ ICB crossings). 
\par Filtering the extreme BS events leads to very similar results to the original BS results, with exactly the same ranking according to AIC and the same three first drivers for LASSO (IMF, $\theta_{bn}$, sunspot number). The IMF intensity (ranked first) LASSO coefficient or AIC score however decreases significantly with respect to the next drivers for the filtered dataset, thus confirming that IMF (or Mach) has a leading role in the occurrence of extreme BS events. The SW dynamic pressure (ranked fourth) coefficient and score also get significantly reduced after filtering, suggesting a significant role - though less important - in the BS expansion events.
\par The same procedure was applied to the ICB dataset. Filtering the most extreme ICB excursions does not impact neither the ranking nor the scores / coefficients of the EUV fluxes, but SW dynamic pressure gets a higher ranking (close to rank number 2) since IMF and Mach number see e.g. their LASSO coefficients decrease by $\sim 40\%$ leading to lower rankings. This suggests that the Mach number and IMF intensity or topology play a significant role in the extreme far ICB events.

\section{Conclusions}
\label{Conclusions} 

\par We analyzed in this paper the physical drivers of the Venusian Bow Shock (BS) and dayside Ion Composition Boundary (ICB) locations. We used a one-dimensional approach based on the extrapolated terminator distance for the BS and the radial distance for the ICB, as well as the crossings lists and data from \citet{persson_2023_database}.
\par The plasma boundaries exhibit a complex dynamical behavior that is influenced by a number of drivers: solar extreme ultraviolet fluxes (that ionize and heat the atmosphere, thus increasing mass loading), the solar wind (SW) dynamic pressure that compresses the induced magnetosphere, the Mach number that in particular drives the BS jump conditions and shapes its flaring, the IMF intensity which adds external magnetic pressure, and finally the IMF orientation, whose angular characteristics eventually drive boundary asymmetries through the clock and cone angles, the $\theta_{bn}$ angle between the IMF and the boundary normal, or the relative clock angle between the IMF and the location of the crossings projected in the terminator plane.
\par Previous works analyzed the direct influence of the boundary drivers or proposed semi-empirical models based on fit functions. However, we showed that many of the possible drivers are correlated to each other, which may induce biases in the interpretation when using classical methods, in particular when analyzing the influence of minor drivers. We thus used three different robust mathematical approaches - partial correlations, Akaike Information Criterion, Least Absolute Shrinkage Selection Operator - in order to find both their direct influence on the boundary locations, as well as their cross-correlations. Using several independent methods, compared with direct classical analysis, allows to provide more robust conclusions. This allowed us to rank the drivers and examine both which drivers have the strongest influence on the boundaries, as well as which drivers have minor or no influence. 
\par The main driver of the Venusian BS location is most probably the magnetosonic Mach number, but in the absence of reliable temperature measurements and due to significant cross correlations, the IMF intensity (or the Alfven Mach number) appears the most significant parameter of influence, with high IMF / low Mach numbers leading to an expanding BS. Then, the BS expands for 2) quasi perpendicular shocks, 3) enhanced solar EUV fluxes and reduced SW dynamic pressures, and possibly for crossings occurring along the IMF direction rather than perpendicular to it (relative clock angle influence). Parametric models of the BS should thus include at least Mach number or IMF intensity, $\theta_{bn}$, as well as EUV fluxes and SW dynamic pressure.
\par The ICB is primarily influenced by solar EUV fluxes through the ionization rates and thermal pressure increase. Then a more reduced influence can be seen from SW parameters that are either IMF intensity and Alfven Mach number together or SW dynamic pressure alone (which shape the SW draping and induced compression of the Venus ionized atmosphere), as well as IMF induced asymmetries (e.g. pole vs equator asymmetry via the relative clock angle) or low cone angle conditions that reduce the magnetic pile-up and thus the external pressure on the ionosphere. Note that the EUV fluxes influence is related to a seasonal influence that takes place on large periods, that can be hidden behind more dynamic drivers (i.e. SW related parameters) if shorter periods of time are considered.
\par The strong cross correlations between the SW parameters make it difficult to identify clearly the influence of asymmetries due to the IMF orientation. Comparing several methods allows to confirm - beyond the strong $\theta_{bn}$ angle influence for the BS - a probable influence of relative clock angle (i.e. pole vs equator asymmetry in the IMF projected frame) for both the BS and the ICB as well as of cone angle for the ICB, while influences of cone and clock angles on the BS are not excluded.


\par A comparison is proposed between the Martian and Venusian bow shock drivers, based on G22 and this study. Overall, both planets share the same main drivers for the BS location, except for a stronger expected relative influence of solar EUV at Mars than at Venus and an apparent stronger influence of relative clock angle (pole vs equator asymmetry) and possibly cone angle at Venus. 
\par Our results suggest that the convective electric field may have less influence on the BS dynamics and shape than expected in previous works, with e.g. a small statistical significance of cone angle influence after taking into account the existing significant cross correlations. \citet{xu_q:2023} observed a small solar wind control (compared to the EUV influence) of the pickup ion plume in the dayside magnetosheath of Venus, suggesting a moderate solar wind control of the planetary mass loading to the solar flow compared to EUV induced ionization, which is in agreement with the reduced impact of cone angle (and thus of the convective electric field) observed in our analysis.
\par Comparing both the ICB and BS distributions as well as their interdistance shows that both boundaries usually share an overall similar dynamics, mostly influenced by the same drivers, but extreme boundary excursions (essentially expansions) may be induced by different conditions. The Venusian magnetosheath size (on average $\sim 70$ ion inertial lengths at the terminator plane) is maximum for strong IMF intensities, quasi-perpendicular shock conditions and strong incident EUV fluxes. A dedicated analysis of the extreme boundary locations reveals that most of the BS extreme events correspond to conditions where several parameters showed severe conditions rather than extreme conditions for a unique driver, though IMF seems to play a key role in the appearance of extreme BS excursions. Extreme ultraviolet fluxes appear as the main origin for extreme ICB expansions, but combined influence of several drivers may also lead to such events.

\section{Open Research}
The datasets used in this study are available in the \url{https://doi.org/10.5281/zenodo.7679678} repository.

\begin{acknowledgments}
This work was supported by the French space agency CNES. We also acknowledge the European Space Agency (ESA) for successfully supporting the VEX mission.
\end{acknowledgments}

\bibliographystyle{aasjournalv7}



\end{document}